\documentclass[fleqn,usenatbib]{mnras}

\usepackage[T1]{fontenc}

\DeclareRobustCommand{\VAN}[3]{#2}
\let\VANthebibliography\thebibliography
\def\thebibliography{\DeclareRobustCommand{\VAN}[3]{##3}\VANthebibliography}

\usepackage{graphicx}	
\usepackage{amsmath}	
\usepackage{amssymb}	

\usepackage{newtxtext,newtxmath}
\DeclareMathOperator{\Tr}{Tr}

\usepackage{float}
\usepackage{pdflscape}

\title[Dynamical age of Tucana-Horologium]{Dynamical age of the Tucana-Horologium young stellar association\thanks{Based on observations collected at the European Southern Observatory under ESO programme 108.2252.001.}}

\author[Galli et al.]{
Phillip A. B. Galli,$^{1}$\thanks{E-mail: phillip.ab.galli@gmail.com}
Núria Miret-Roig,$^{2}$
Hervé Bouy,$^{3}$
Javier Olivares$^{4}$
and David Barrado$^{5}$
\\
$^{1}$ Núcleo de Astrofísica Teórica, Universidade Cidade de São Paulo, R. Galvão Bueno 868, Liberdade, 01506-000 São Paulo, SP, Brazil.\\
$^{2}$University of Vienna, Department of Astrophysics, Türkenschanzstraße 17, 1180 Wien, Austria.\\
$^{3}$Laboratoire d’astrophysique de Bordeaux, Univ. Bordeaux, CNRS, B18N, allée Geoffroy Saint-Hilaire, 33615 Pessac, France.\\
$^{4}$Depto. de Inteligencia Artificial, UNED, Juan del Rosal, 16, 28040 Madrid, Spain.\\
$^{5}$Centro de Astrobiología (CSIC-INTA), ESAC Campus, Camino Bajo del Castillo s/n, 28692 Villanueva de la Cañada, Madrid, Spain.
}

\date{Accepted XXX. Received YYY; in original form ZZZ}

\pubyear{2015}

\begin{document}
\label{firstpage}
\pagerange{\pageref{firstpage}--\pageref{lastpage}}
\maketitle

\begin{abstract}
The Tucana-Horologium association is one of the closest young stellar groups to the Sun and despite the close proximity its age is still debated in the literature. We take advantage of the state-of-the-art astrometry delivered by the third data release of the \textit{Gaia} space mission combined with precise radial velocity measurements obtained from high-resolution spectroscopy to investigate the dynamical age of the association. We perform an extensive traceback analysis using a combination of different samples of cluster members, metrics to evaluate the minimum size of the association in the past and models for the galactic potential to integrate the stellar orbits back in time. The dynamical age of $38.5^{+1.6}_{-8.0}$~Myr that we derive in this paper is consistent with the various age estimates obtained from isochrone fitting in the literature (30-50~Myr) and reconciles, for the first time, the dynamical age of the Tucana-Horologium association with the age obtained from lithium depletion ($\sim40$~Myr). Our results are independent from stellar models and represent one more step towards constructing a self-consistent age scale for the young stellar groups of the Solar neighbourhood based on the 3D space motion of the stars.
\end{abstract}

\begin{keywords}
Galaxy: kinematics and dynamics - stars: kinematics and dynamics - open clusters and associations: individual: Tucana-Horologium association - solar neighbourhood - stars: formation.
\end{keywords}

\section{Introduction}

The young stellar associations within 100~pc from the Sun are valuable benchmarks to study the evolution and dispersal of young stars in the field. The close proximity and young age has also made them ideal targets to search for planetary-mass companions around young stars \citep[see e.g.][]{Chauvin2004,Lagrange2010,Newton2019} and studies of circumstellar disks that depend on angular resolution \citep[see e.g.][]{Smith1984,Golimowski2006,Boccaletti2009}. These stellar groups are mostly sparse and contain a few dozen to hundreds of stars spread over large portions of the sky \citep[see e.g.][]{Torres2008}. It is implicitly assumed that the members of a stellar association are coeval and co-moving as they were born from the same star formation episode. So, while the determination of the fundamental parameters of individual stars e.g. age and spatial velocity is sometimes challenging, the study of a large ensemble of stars that belong to the same association makes it possible to infer the group properties with relatively good precision \citep[see e.g.][]{Gagne2018a,Bell2015}. 

The Tucana-Horologium association (hereafter, Tuc-Hor) located at about 46~pc \citep{Gagne2018a} is one of the most prominent young stellar associations of the Solar neighbourhood. Its members were identified separately as two different stellar groups, the Tucana and Horologium associations \citep{Torres2000,Zuckerman2000}, but later recognized to form one single population of young stars \citep{Zuckerman2001}. One intriguing point about the Tuc-Hor association is that its age is still debated in the literature despite the progress that has been made in recent years to characterize the stellar population. The various age estimates obtained for Tuc-Hor over time range from about 27~Myr to 53~Myr \citep{Torres2000,Zuckerman2000,Zuckerman2004,Mentuch2008,Kraus2014,Bell2015,Kerr2022}. One reason to explain the relatively broad range of age results published in the literature is of course the different samples of Tuc-Hor stars and data employed in each study. However, it is also important to mention that most of these age estimates were obtained from methods (e.g. isochrone fitting) that rely on the choice of evolutionary models which most likely explain the different results that have been produced. A summary of the different age-dating methods for young stars, their strengths and weaknesses  can be found in \citet{Soderblom2010}.

One alternative to compute stellar ages that is independent from stellar models consists in deriving the dynamical age of the association. This method assumes that the members of a stellar association formed together and were more concentrated in the past. Thus, by tracing the stellar positions back in time one can determine the time when the association reached its minimum size in the past and infer the dynamical age of the group \citep[see e.g.][]{delaReza2006,Ducourant2014,Miret-Roig2018}. Despite the simplicity of the method, it relies on precise information on the 3D spatial velocity of the stars which is required to trace the stellar orbits back in time. This challenge can be overcome nowadays by using the state-of-the-art astrometry delivered by the third data release of the \textit{Gaia} space mission \citep[Gaia-DR3,][]{GaiaDR3} combined with ground-based high-resolution spectroscopy to measure precise radial velocities as we explain in this study.

Our team has recently derived the dynamical age of the $\beta$~Pictoris moving group (hereafter, $\beta$~Pic) and confirmed, for the first time, that the dynamical age is consistent with the age obtained from other age-dating methods \citep[see][]{Miret-Roig2020}. In that study we demonstrated that the use of a homogeneous and precise dataset of radial velocities, a careful selection of the most likely group members, and a robust statistical approach to evaluate the size of the association are crucial points in the traceback analysis to accurately infer the dynamical age of the association. The success of this pilot study encouraged us to extend the analysis to other young stellar associations with the final objective of constructing a self-consistent age scale for the young stellar groups in the Solar neighbourhood.

This paper is one in a series dedicated to investigate the dynamical age of young stellar groups and it will focus on the Tuc-Hor association. It is structured as follows. In Section~\ref{section2} we introduce the sample of stars and data that will be used in this study. In particular, we describe our methodology to derive precise radial velocities by combining high-resolution archival spectra with our own observations, and present our selection criteria to define a clean sample of the most likely kinematic members of the association. Then, we present in Section~\ref{section3} our strategy to infer the dynamical age of the association using different metrics to estimate the size of the association, models for the Galactic potential and samples of stars in the traceback analysis. In Section~\ref{section4} we compare our result for the dynamical age of Tuc-Hor with other age estimates derived from different methods in the literature. Finally, we summarize our results in Section~\ref{section5}.

\section{Sample and Data}\label{section2}

\subsection{Census of cluster members}\label{section2.1}

Our initial sample of Tuc-Hor stars is based on the lists of cluster members delivered by the BANYAN project \citep{Gagne2018a,Gagne2018b,Gagne2018c}. The BANYAN classification tool infers membership probabilities of stars to the nearest young stellar associations within 150~pc based on their position and spatial velocity. The various lists of candidate members identified in that project constitute therefore a valuable benchmark for the study of the young local associations. This compilation of Tuc-Hor cluster members contains 94~stars after removing the sources in common among the several lists that have been updated over time. The components of binary and multiple systems that have been resolved in previous studies are counted as independent entries in this sample. No attempt has been made in this paper to search for new cluster members of the Tuc-Hor association which clearly goes beyond the scope of our study. On the contrary, we have refined the census of cluster members in light of the new astrometric and spectroscopic data that is currently available by selecting the most likely kinematic members of the association before performing the traceback analysis as explained below. 

\subsection{Proper motions and parallaxes}\label{section2.2}

The Gaia-DR3 catalogue represents a major advance in the series of data releases from the \textit{Gaia} space mission \citep{Gaia} featuring a wealth of data products. The more precise astrometry and radial velocities included in Gaia-DR3 are of particular interest in our study to compute the 3D spatial velocity of the stars. So, we cross-matched our initial sample of Tuc-Hor stars with the Gaia-DR3 catalogue and retrieved the proper motions and parallaxes for 90 stars. The four missing sources without astrometric solution correspond to the components of the HD~20121~AB (Gaia DR3 4850405974392619648) and HD~207964~AB (Gaia~DR3~6408955426068786688) systems which were not resolved by the \textit{Gaia} satellite. 

The median precision in proper motion (right ascension and declination) and parallax that we obtain with our sample are 0.020~mas/yr, 0.022~mas/yr and 0.019~mas, respectively. This translates into a typical precision of about 0.022~km/s in the 2D tangential velocities of the stars.


\subsection{Radial velocities}\label{section2.3}

After the launch of the \textit{Gaia} satellite, the scarcity and precision of the radial velocity data are currently the two main limitations to compute the 3D space motion of the stars and perform a traceback analysis \citep[see][]{Miret-Roig2018,Miret-Roig2020}. Although Gaia-DR3 provides information on the radial velocity of the stars, these measurements are restricted to the brightest sources in the catalogue ($G_{RVS}<14$~mag) and they often exhibit large uncertainties as discussed below. Previous studies delivered the radial velocity for a number of stars in our sample \citep[see e.g.][]{Torres2006,Gontcharov2006,Holmberg2007,Kraus2014}. However, a compilation of the published results in the literature based on different methodologies to compute the radial velocity and its uncertainty results in a heterogenous data set. We therefore resort to download the spectra from public archives and derive the radial velocities ourselves to ensure that the same methodology is used to estimate the radial velocity for all stars in our sample and produce consistent results. In addition, we complemented the archival data with our own observations as we explain below. 

\subsubsection{Spectra from public archives}\label{section2.3.1}

We searched for high-resolution spectra ($R\gtrsim40\,000$) of the stars in our sample in public archives and found a total of 789 spectra in the European Southern Observatory (ESO) and \textit{Observatoire Haute-Provence} (OHP) databases. The spectra were reduced with the pipeline that is available for each instrument. In Table~\ref{tab_archival_spectra} we present the number of spectra retrieved for each instrument. In most cases we found multiple spectra for the same star (with the same or various instruments) which explains the large number of spectra downloaded from the archives compared to the number of stars in our sample.

\begin{table}
\centering
\caption{Properties of the spectra downloaded from public archives. We provide for each instrument (spectrograph) the maximum resolving power, spectral range, and number of spectra downloaded from the archives.}
\label{tab_archival_spectra}
\begin{tabular}{lccc}
\hline\hline
Instrument&$R$&$\Delta\lambda$&number of spectra\\
&&(nm)&\\
\hline\hline
ESO/FEROS&48\,000&350-920&232\\
ESO/HARPS&115\,000&378-691&451\\
ESO/UVES&110\,000&300-1100&98\\
OHP/SOPHIE&75\,000&387-694&8\\
\hline\hline
\end{tabular}
\end{table}

\subsubsection{Spectra from our observations}\label{section2.3.2}

We performed our own observations with the Ultraviolet and Visual Echelle Spectrograph \citep[UVES,][]{Dekker2000} operated at ESO~(Paranal, Chile) to increase the number of stars in our sample with radial velocity information (programme identifier: 108.2252). We targeted a sample of 17 stars in the range of magnitude from about $V=13$ to $V=16$~mag. This is a critical magnitude (and colour) regime that was not covered in previous studies. The observations were performed in service mode from September to December 2021. We used the UVES red arm with the 600~nm standard setting (spectral range from about 500 to 705~nm) and the 1\arcsec slit yielding a resolving power of $R\simeq42\,000$. The exposure times ranged from about 7 to 30~min depending on the magnitude of the target. The collected spectra were reduced with the UVES pipeline and they have mostly a signal to noise ratio (SNR) of about 20-30. 

\subsubsection{Radial velocity determination}\label{section2.3.3}

We computed the radial velocity of the stars using iSpec \citep{BlancoCuaresma2014,BlancoCuaresma2019} as explained below. First, we cross-correlated the spectra with masks of different spectral types (A0, F0, G2, K0, K5, and M5) and computed the cross-correlation function (CCF). The radial velocity given by iSpec is computed by fitting a second order polynomial near the peak of the CCF and the radial velocity uncertainty is estimated following the procedure outlined in Sect.~2.3 of \citet{Zucker2003}. We report for each star the radial velocity value derived from the closest mask to the spectral type of the star. However, to account for the observed fluctuation in the results obtained with different masks we computed the radial velocity scatter obtained from three different masks (the closest mask to the spectral type of the star, one before and one after) and added this number in quadrature to the formal uncertainty given by iSpec. The later step is likely to overestimate the radial velocity uncertainty for some stars and represents therefore an upper limit of the uncertainty that is valid for the purposes of this work. We inspected the CCFs individually and discarded the radial velocity results that derived from a poor fit to the CCF that often arise due to a low SNR of the spectrum or a mismatch between the spectral type of the star and the adopted mask. 

Table~\ref{tab_RVs} lists the individual radial velocity measurements obtained for each star in different epochs. We proceed as follows to combine the various radial velocity measurements for the (single) stars with multiple spectra. We use each radial velocity solution (derived from a single spectrum of the same star) to generate 1\,000 synthetic measurements from a Gaussian distribution where the mean and variance correspond to the radial velocity and its uncertainty, respectively. The joint distribution of the ensemble of synthetic radial velocity measurements (obtained after combing the results from multiple spectra) is used to estimate the final parameters of the stars. We take the mean of the joint distribution of radial velocity measurements as our final radial velocity estimate and compute the uncertainty from the 16\% and 84\% percentiles of the distribution. The final radial velocities obtained in this work for each star are given in Table~\ref{tab_data}. By combining the archival data with our own observations we were able to derive the radial velocity for 40~stars in our sample. Our observations performed with the UVES spectrograph represent 28\% of the stars with radial velocity result and they are the faintest sources in the sample as explained before. 

\begin{figure*}
\begin{center}
\includegraphics[width=0.49\textwidth]{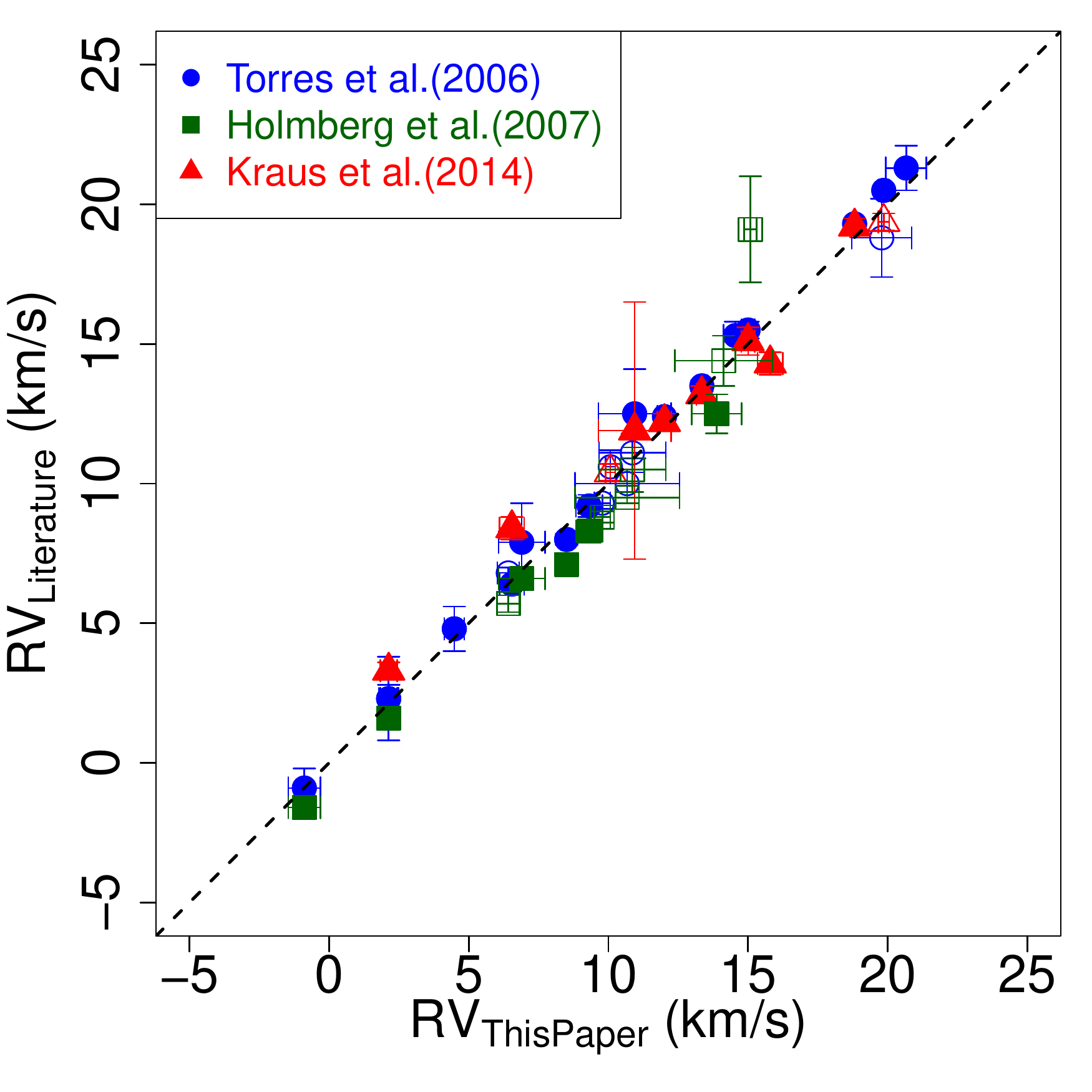}
\includegraphics[width=0.49\textwidth]{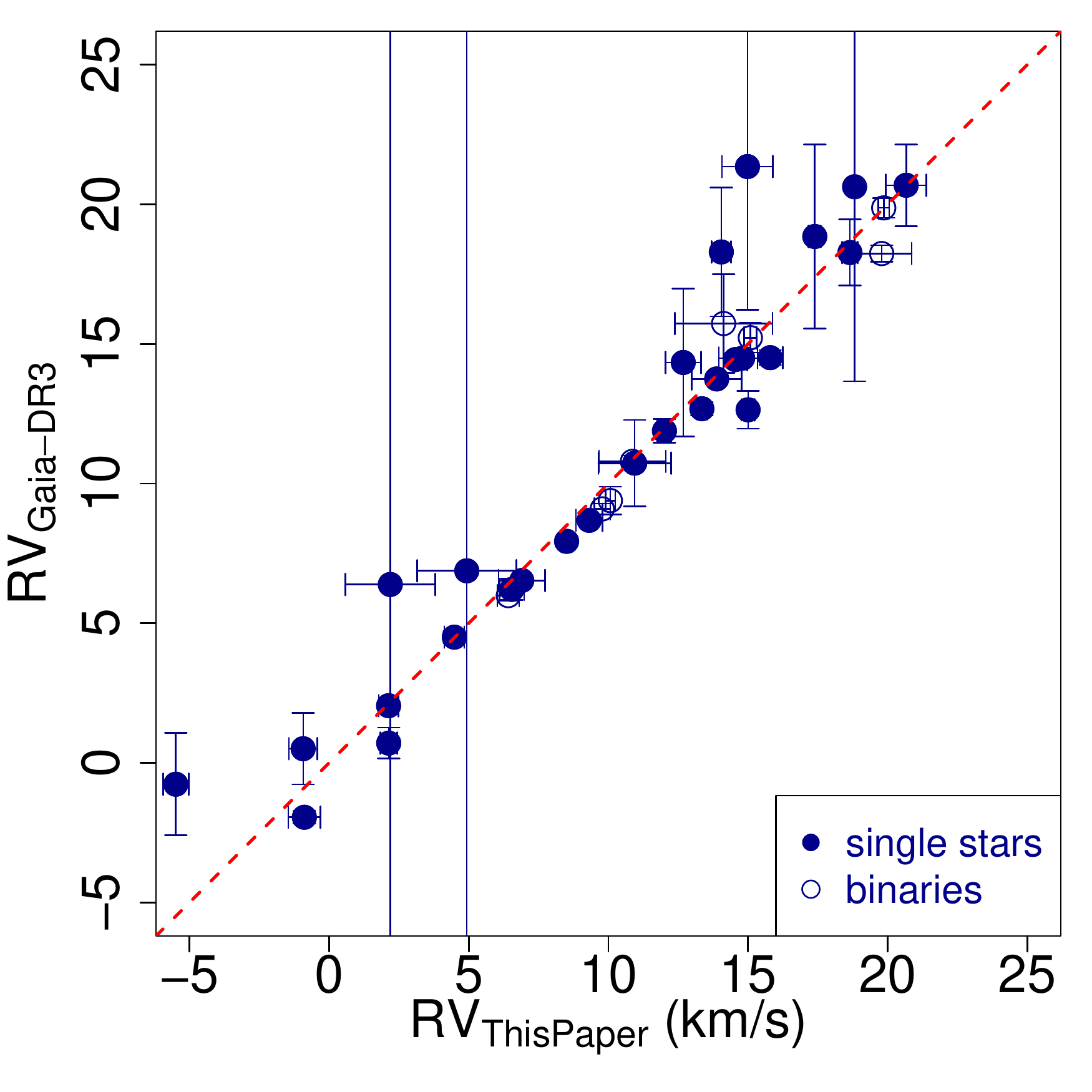}
\caption{Comparison of the radial velocities derived in this paper with several studies (\textit{left panel}) and the Gaia-DR3 catalogue (\textit{right panel}). Filled and open symbols correspond to single stars and binaries, respectively. The dashed line in the plots indicates perfect correlation between the measurements. }
\label{fig_RV_comp}
\end{center}
\end{figure*}

The median radial velocity uncertainty of our sample is 0.4~km/s. We now compare the quality of our radial velocity measurements with respect to published results. In Table~\ref{tab_RV_comp} we compare our radial velocities with the ones given by \citet{Torres2006}, \citet{Holmberg2007}, \citet{Kraus2014}, and the Gaia-DR3 catalogue. We note that the root mean square error (RMSE) of the radial velocities is higher for the comparison with the Gaia-DR3 catalogue which includes many sources with poor radial velocity precision (see also Sect.~\ref{section2.4}). However, as illustrated in Figure~\ref{fig_RV_comp} the radial velocities derived in this paper are overall in good agreement with published results and our measurements are often more precise. Thus, by computing ourselves the radial velocities from archival data we were able to deliver a homogenous sample of precise radial velocities and provide this information for a larger number of stars from our sample. 

\begin{table}
\centering
\caption{Comparison of radial velocity results. We provide for each study that we use in our comparison the number of stars in common (the number in parentheses refers to the number of single stars), the mean difference in radial velocity (in the sense, this paper ``minus’’ other study), and the root mean square error (RMSE).
\label{tab_RV_comp}}
\begin{tabular}{lccc}
\hline\hline
Catalogue&Number of stars&<$\Delta RV$>&RMSE\\
&&(km/s)&(km/s)\\
\hline\hline
\citet{Torres2006}&22(16)&$-0.2\pm0.1$&1.1\\
\citet{Holmberg2007}&12(6)&$+0.4\pm0.4$&1.2\\
\citet{Kraus2014}&10(8)&$-0.3\pm0.3$&0.9\\
Gaia-DR3&35(27)&$-0.5\pm0.3$&2.7\\
\hline\hline
\end{tabular}
\end{table}

\subsection{Sample selection}\label{section2.4}
The radial velocities derived by ourselves and the Gaia-DR3 catalogue are currently the main sources of radial velocity information for the stars in our sample. The Gaia-DR3 catalogue provides radial velocity data for 60~stars of our initial sample, but many of these sources have large uncertainties and they cannot be used in the traceback analysis. We therefore restrict the Gaia-DR3 radial velocities to the sources with a radial velocity precision $<2$~km/s which roughly corresponds to the least precise radial velocity measurement derived by ourselves (see Table~\ref{tab_data}). This reduces the number of sources in the Gaia-DR3 catalogue with useful data to 40~stars.

We obtain the largest sample of Tuc-Hor stars with available radial velocity information by combining our radial velocity measurements with the ones given by Gaia-DR3. This yields a sample of 52~ stars with complete 6D data (5D astrometry and radial velocity).  For the sources with radial velocity data available in the two projects we use the ones derived by ourselves. Alternatively, we also investigate the dynamical age of the association using two control samples. The first control sample (hereafter, CS1) uses only the radial velocities derived by ourselves combined with the 5D astrometry from the Gaia-DR3 catalogue. CS1 contains 40~stars (see Sect.~\ref{section2.3.3}). The second control sample (hereafter, CS2) takes only the sources in the Gaia-DR3 catalogue with complete 6D data (with a radial velocity precision better than 2~km/s as explained above) and it contains 40~stars. The same number of stars in CS1 and CS2 is merely a coincidence, because the two samples are obviously not identical. 

\begin{figure*}
\begin{center}
\includegraphics[width=0.33\textwidth]{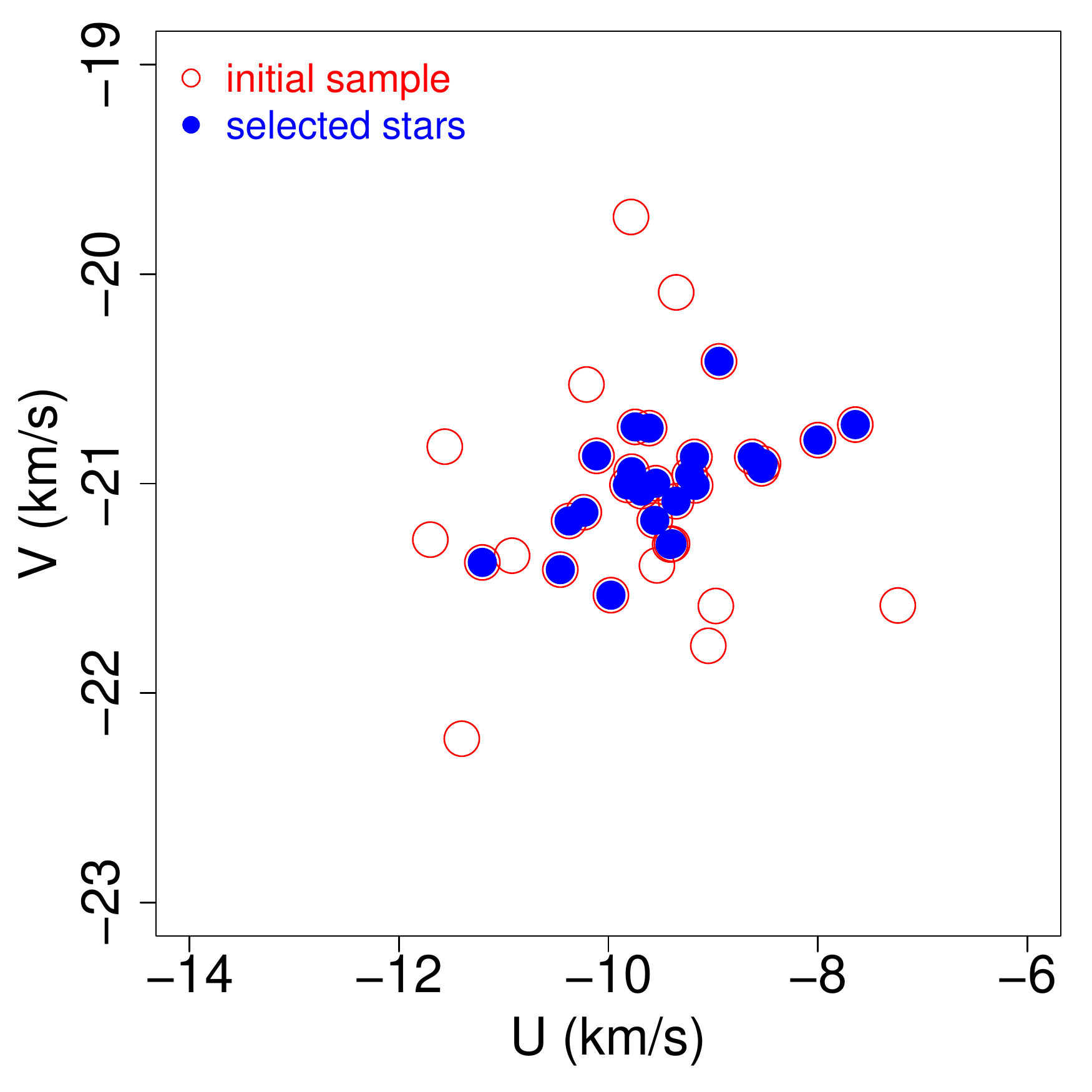}
\includegraphics[width=0.33\textwidth]{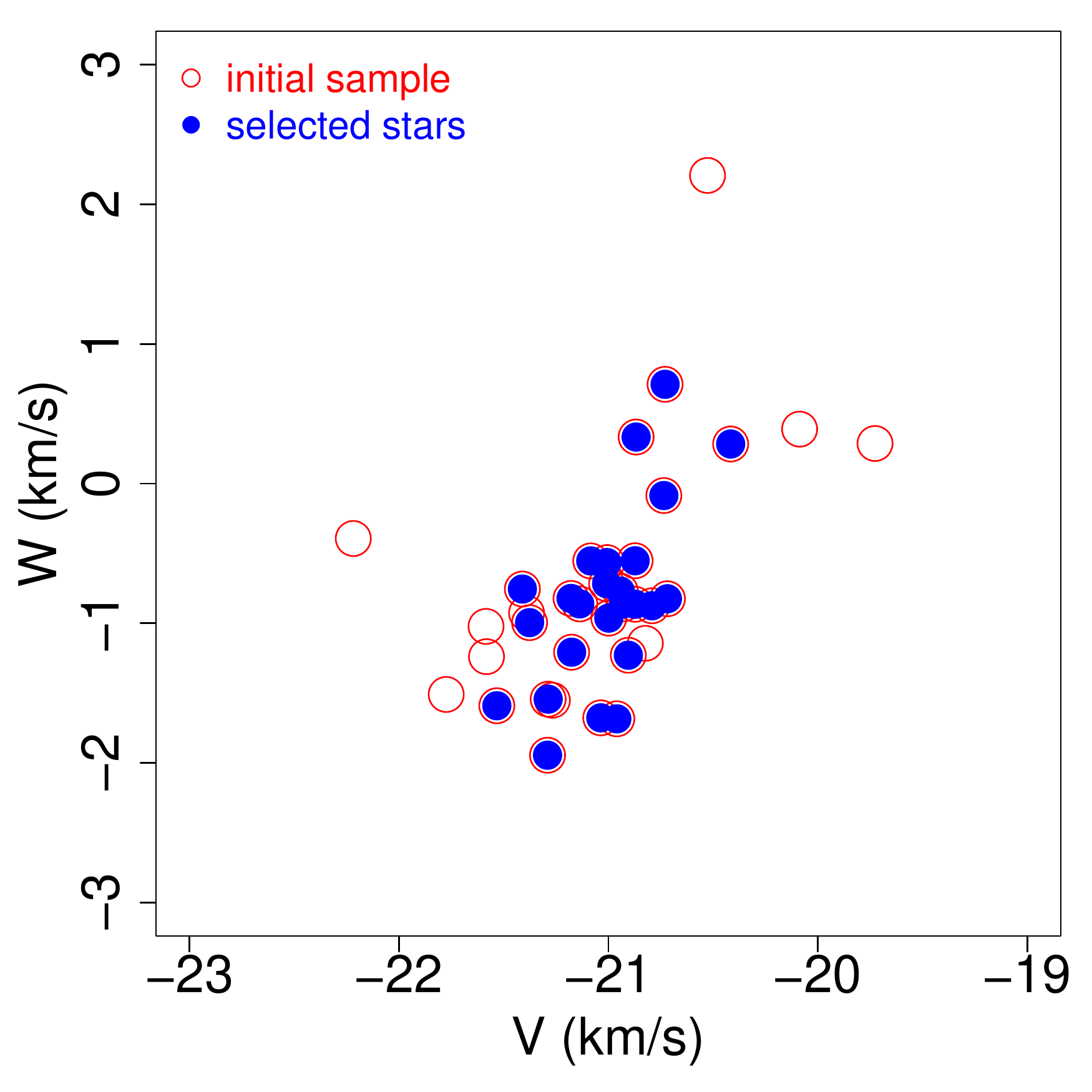}
\includegraphics[width=0.33\textwidth]{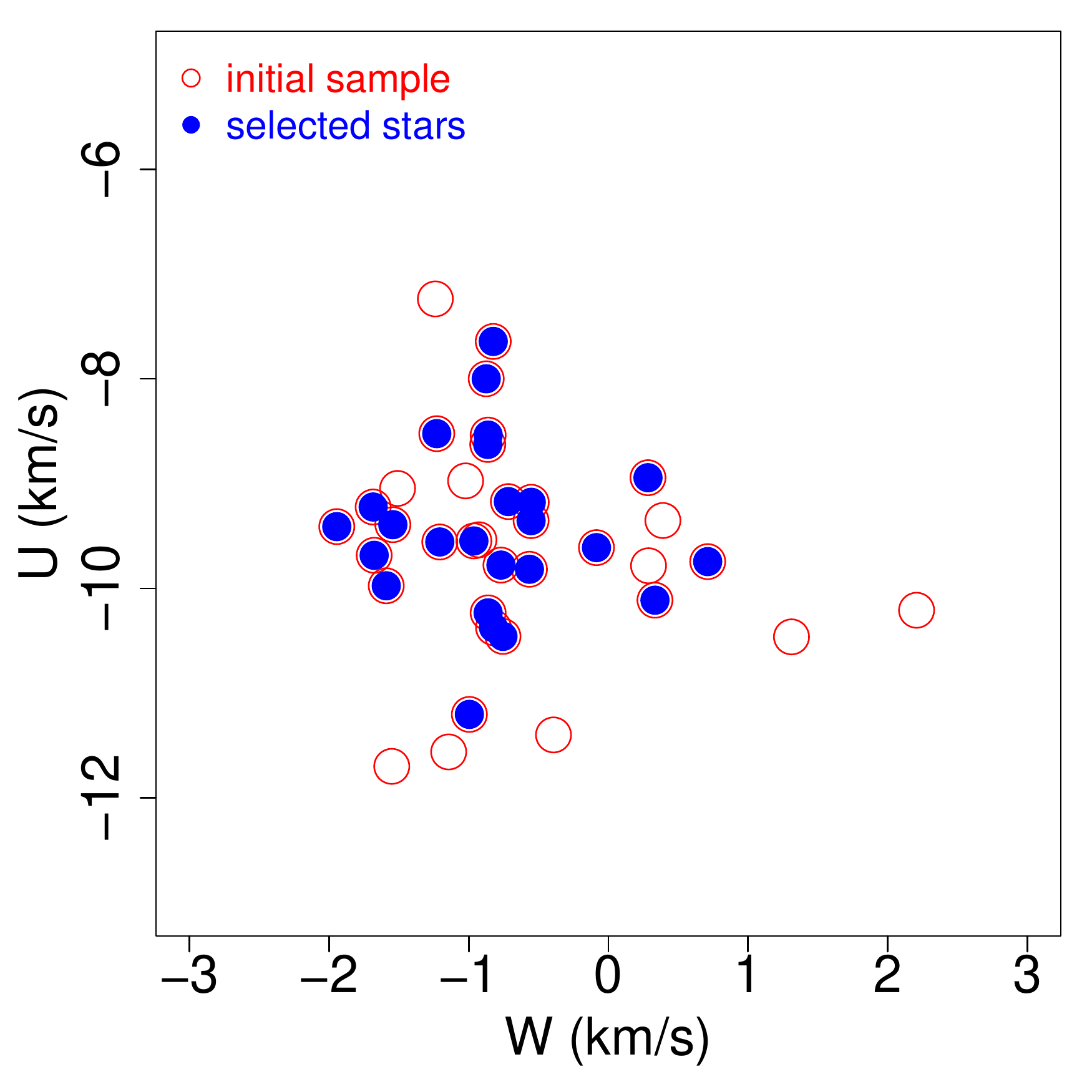}
\caption{Sample selection in the 3D space of velocities based on robust distances using the MCD estimator.}
\label{fig_MCD}
\end{center}
\end{figure*}

The next step in our analysis consists in filtering the stars that will be used in the traceback analysis. We proceed as follows to select a clean sample of cluster members with reliable data. First, we remove known binaries and multiple systems from our sample because the variable radial velocity of these sources can affect the traceback analysis. Second, we compute the 3D spatial velocity of the stars and remove potential outliers in the space of velocities under the assumption that members of a stellar association share the same space motion. We computed the $UVW$ Galactic velocity of the stars using the same reference system defined by \citet{Johnson1987} where $X$ points to the Galactic centre, $Y$ points in the direction of the Galactic rotation and $Z$ points to the Galactic north pole. Then, to identify potential outliers in the 3D velocity space we compute robust distances (RD) which are given by
\begin{equation}
RD(\mathbf{x})=\sqrt{(\mathbf{x}-\boldsymbol{\mu})^{t}\boldsymbol{\Sigma}^{-1}(\mathbf{x}-\boldsymbol{\mu})}\, ,
\end{equation} 
where $\boldsymbol{\mu}$ and $\boldsymbol{\Sigma}$ denote the multivariate location and covariance matrix that are obtained from the minimum covariance determinant \citep[MCD,][]{Rousseeuw1999} method. We used a tolerance ellipse of 95\% to identify potential outliers in the velocity space. Figure~\ref{fig_MCD} illustrates the outcome of this selection procedure for our sample. We present in Table~\ref{tab_samples} the number of sources that have been retained in our sample, CS1 and CS2 after applying our selection criteria. 

As a by-product of our analysis we compute the mean space motion of $(U,V,W)=(-9.7,-21.0,-0.9)\pm(0.1,0.1,0.1)$~km/s for the Tuc-Hor association. This result is consistent with previous estimates for the 3D space motion of Tuc-Hor given in the literature \citep[see e.g.][]{Torres2008,Kraus2014}. The one-dimensional velocity dispersion that we find from the standard deviation of each velocity component is $(\sigma_{U},\sigma_{V},\sigma_{W})=(0.5,0.2,0.7)$~km/s and the median uncertainty in the $UVW$ velocity of the stars is 0.1, 0.2, and 0.3~km/s, respectively. This confirms that Tuc-Hor stars are co-moving with a relatively low intrinsic velocity dispersion.

\begin{table*}
\centering
\caption{Number of stars in our sample and control samples (CS) after applying the selection criteria.}
\label{tab_samples}
\begin{tabular}{lccc}
\hline\hline
&CS1&CS2&Sample\\
\hline\hline
Sample with radial velocity data ($\sigma_{RV}<2$~km/s)&40 stars& 40 stars& 52 stars\\
Sample of single stars with 6D data&31 stars& 26 stars& 37 stars\\
Sample of selected single stars in the velocity space&24 stars& 15 stars& 25 stars\\
Final sample confirmed by orbital analysis&20 stars& 14 stars& 21 stars\\
\hline\hline
\end{tabular}
\end{table*}

\section{Traceback Analysis}\label{section3}

In this section we present our strategy to infer the dynamical age of the Tuc-Hor association. We define the dynamical age of a stellar association as the time in the past when the stars were more concentrated in space and the group had its minimum size. The methodology that is used to estimate the size of the association is one critical point in the traceback analysis. Here we investigate the size of the association with three different metrics that have been successfully employed in our previous study of the $\beta$~Pic association \citep[][]{Miret-Roig2020}. They are defined as follows. 

First, we use the squared root of the elements in the main diagonal of the covariance matrix in the $X$,$Y$,$Z$ directions (hereafter, $S_{X}$, $S_{Y}$, $S_{Z}$) to define the size of the association in each direction. Second, we estimate the size of the association as
\begin{equation}
S_{TCM}=\left[\frac{\Tr(\Sigma)}{3}\right]^{1/2}
\end{equation}
where the trace of the covariance matrix $\Sigma$ returns a measure of the total variance (i.e. the sum of its eigenvalues) and the multiplicative factor of 1/3 yields the arithmetic mean of the variances in all three directions. Third, we estimate the size of the association from the determinant of the covariance matrix defined as
\begin{equation}
S_{DCM}=[\det(\Sigma)]^{1/6}
\end{equation}
given that the determinant of the covariance matrix is related to the volume of the association. It is important to mention that the covariance matrix $\Sigma$ that is used in the definition of the size estimators with all three metrics is the covariance matrix computed from the MCD method (see Sect.~\ref{section2.4}). The resulting covariance matrix provides a more robust estimator of the multivariate location and scatter of the stars that is less sensitive to the presence of outliers in the sample compared to the empirical covariance matrix estimation \citep[see also][]{Miret-Roig2020}. We used the Python implementation of the MCD method from the \textit{scikit-learn} package \citep{scikit-learn} to compute the different size estimators in the traceback analysis. 

\begin{figure*}
\begin{center}
\includegraphics[width=0.7\textwidth]{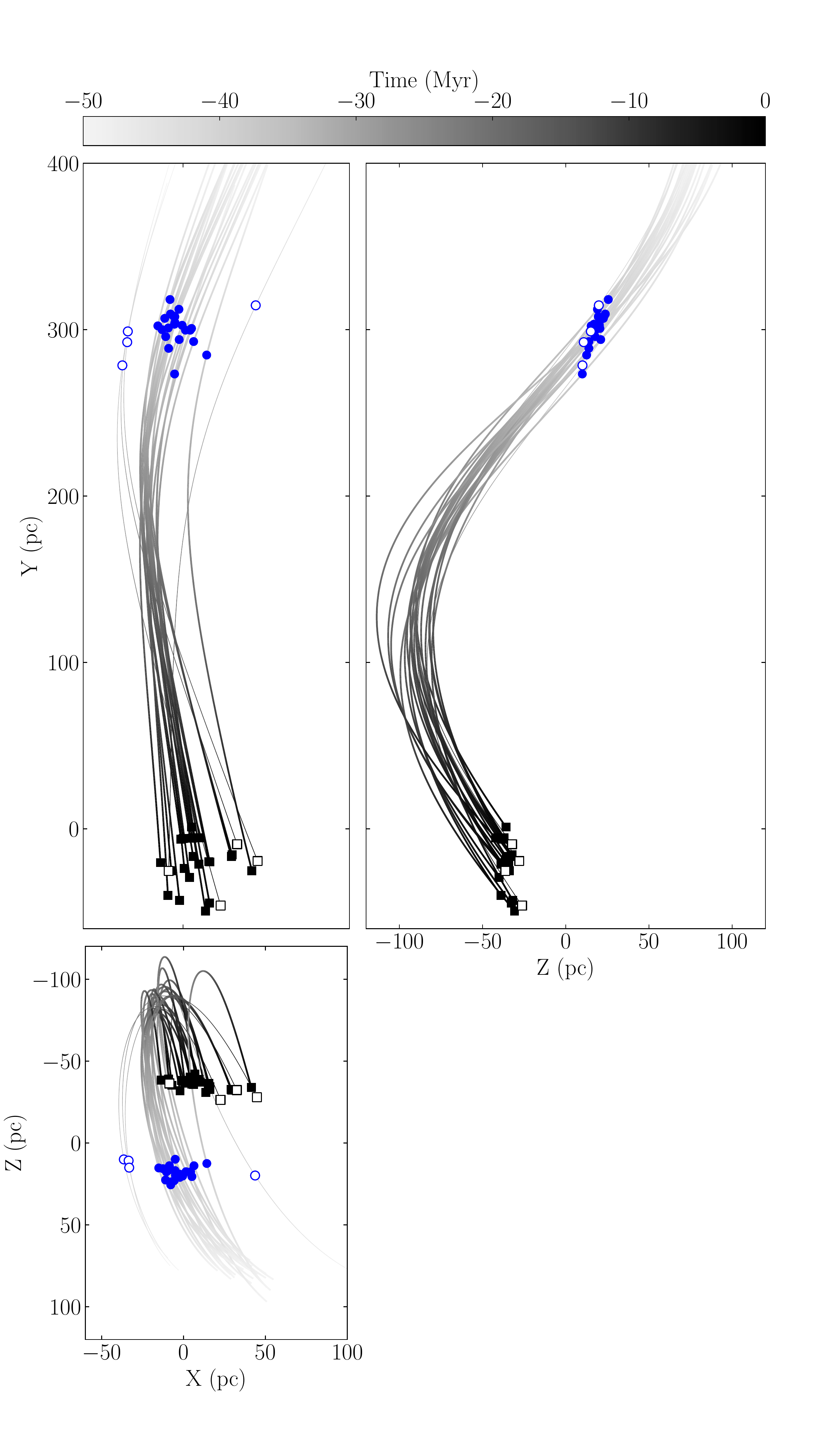}
\caption{2D projection of the stellar orbits from the 25 stars in our sample before removing the interlopers. The black squares indicate the present-day position of the stars ($t=$~0) and the blue circles denote the stellar positions at birth time ($t=-38.5^{+1.6}_{-8.0}$~Myr). The orbits are colour-coded based on the traceback time. Open symbols denote the four stars that exhibit diverging orbits with respect to the rest of the group (see Section~\ref{section3}).}
\label{fig_2d_orbits}
\end{center}
\end{figure*}

We integrate the stellar orbits back in time from $t=0$ (present-day position) to $t=-80$~Myr in steps of 0.1~Myr using the same reference system as defined in Sect.~\ref{section2.4} and the solar motion of $(U_{\odot},V_{\odot},W_{\odot})=(11.10,12.24,7.25)$~km/s \citep{Schoenrich2010}. Our traceback analysis uses the Milky Way's galactic potential (hereafter, MWPotential2014) included in the \textit{galpy} Python library for galactic dynamics \citep{Bovy2015} to integrate the stellar orbits. A visual inspection of the stellar orbits reveals the existence of a few stars in the sample with diverging orbits. In Figure ~\ref{fig_2d_orbits} we illustrate the 2D projection of the stellar orbits in our sample. Four stars (Gaia DR3 4848412525451859328, Gaia DR3 5194203331750283392, Gaia DR3 6445465087626049920 and Gaia DR3 6463782435948431616) are clearly diverging from the rest of the group in the past although their present-day position and space motion are consistent with the other stars. The presence of these interlopers in the sample does not significantly affect the dynamical age of the association, but increases the uncertainty of our solution. We have therefore decided to remove these sources from the analysis. This yields a final sample of 21~stars (see Table~\ref{tab_samples}) that is used here to investigate the dynamical age of the Tuc-Hor association. 

\begin{figure*}
\begin{center}
\includegraphics[width=1.0\textwidth]{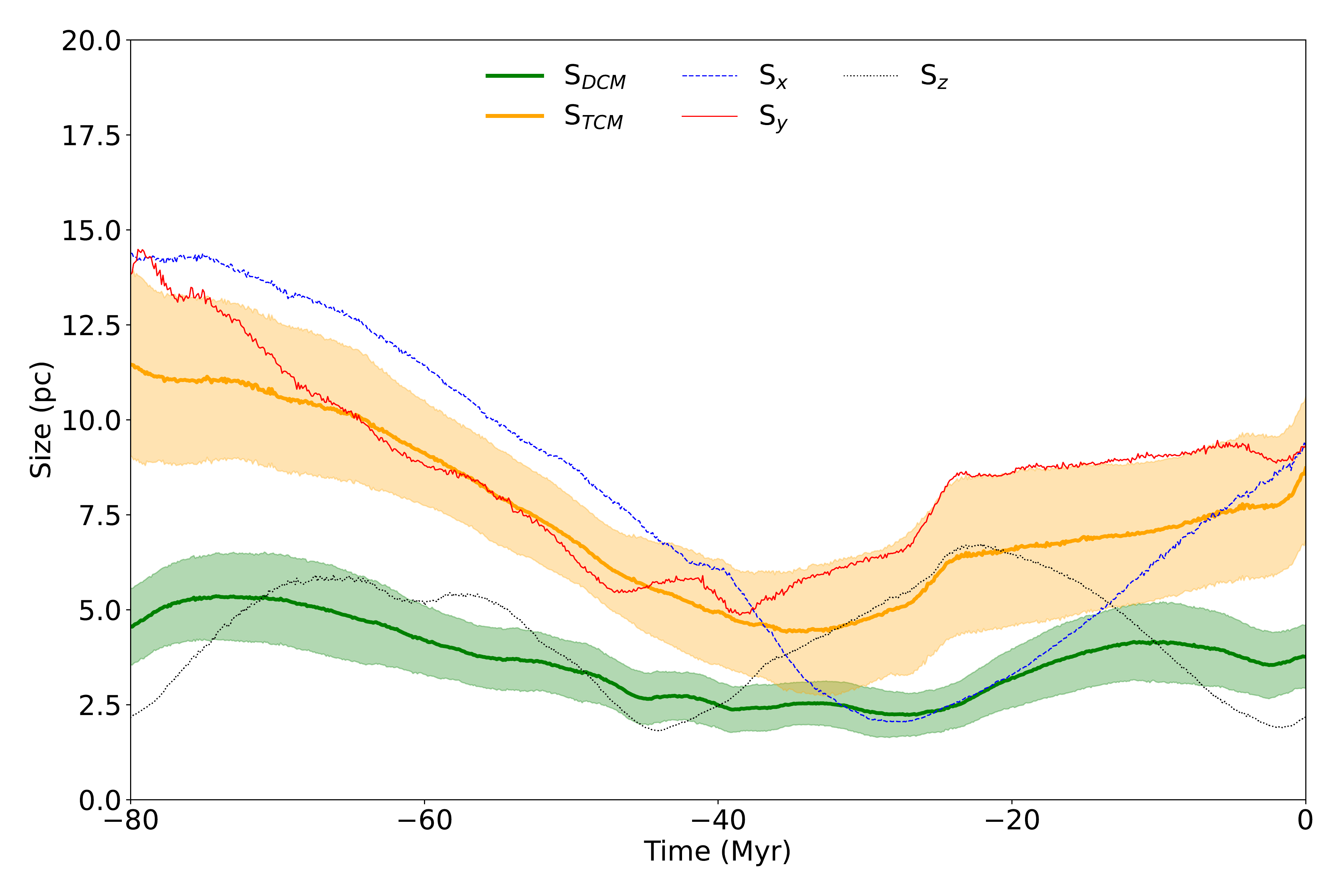}
\caption{Evolution of the size of the association as a function of time using different estimators. The shaded regions represent the 1$\sigma$ uncertainties in the size of the association obtained from 1\,000 bootstrap iterations and the lines indicate the median values obtained for each method.}
\label{fig_size_time}
\end{center}
\end{figure*}

Figure~\ref{fig_size_time} shows the size of the association as a function of time using the different metrics employed to define the size of the association. We note that the minimum size of the association inferred from the elements in the main diagonal of the covariance matrix ($S_{X},S_{Y},S_{Z}$) depends on the direction that is considered in the analysis. The $S_{X}$ size estimator points to a dynamical age of about 29~Myr while the $S_{Z}$ estimator indicates an age value of about 44~Myr. The result obtained from $S_{Y}$ size estimator which is measured in the $Y$ direction (in the direction of Galactic rotation) returns an age of about 39~Myr and is more consistent with the results obtained from $S_{Z}$. The problem of measuring different values for the size of the association in each direction is overcome by using the $S_{TCM}$ estimator based on the trace of the covariance matrix which averages the contribution of all three directions. The minimum size of the association inferred from the $S_{TCM}$ estimator occurs at about 39~Myr and it is more consistent with the results obtained with $S_{Y}$ and $S_{Z}$. The component that contributes most to the size of the association inferred from the $S_{TCM}$ estimator is in the $Y$ direction given that the minimum size of the association obtained from $S_{Y}$ ($\sim$5~pc) is larger than the values obtained from $S_{X}$ ($\sim$2~pc) and $S_{Z}$ ($\sim$ 2~pc). On the other hand, the size of the association inferred from the $S_{DCM}$ based on the determinant of the covariance matrix defines a plateau over a time interval of about 20~Myr which roughly extends from the minimum of the curves obtained from the $S_{X}$ and $S_{Z}$ estimators. This makes the locus of the minimum size (and therefore the dynamical age) of the association very uncertain in contrast to the results obtained from the other methods. We have therefore decided to use the $S_{TCM}$ estimator to infer the size of the association which takes the dispersion of the stars in all three directions into account and provides a more robust estimate of the dynamical age of the group. 

We present in Table~\ref{tab_age} our results for the dynamical age of the association obtained from the different samples and size estimators. We follow the same procedure of \citet{Miret-Roig2020} to report our results for the dynamical age as explained below. First, we perform 1\,000 bootstrap repetitions using random samples of the group members selected in our analysis and compute the dynamical age for each sample generated in this procedure. Then, we take the mode of the age distribution obtained from the bootstrap iterations and the 68\% highest-density interval to determine the dynamical age of the association with its uncertainty. Our results in Table~\ref{tab_age} show that the dynamical ages obtained from different samples using the same size estimator agree within 3~Myr and they are compatible among themselves within the reported uncertainties. In particular, we note that the results obtained with our final sample of stars and CS1 are consistent within 1~Myr. This can be explained by the large overlap of stars between these two samples. Although the initial input catalogue that is used to construct our sample combines the radial velocities derived in this paper with the Gaia-DR3 catalogue, the final sample that is obtained after applying our selection criteria (see Sect.~\ref{section2.4}) produces a sample where most stars are in common with CS1.
We also note that the age results obtained with CS2 have larger uncertainties as compared to CS1 (despite the similar precision in the radial velocity of the sources in common) which comes from the smaller number of stars that is used in the traceback analysis.

\begin{table*}
\centering
\caption{Dynamical age of the association derived from different size estimators.}
\label{tab_age}
\begin{tabular}{lcccccc}
\hline\hline
&Stars&$S_{X}$ & $S_{Y}$ & $S_{Z}$ & $S_{TCM}$ & $S_{DCM}$\\
&&(Myr)&(Myr)&(Myr)&(Myr)&(Myr)\\
\hline\hline
\vspace{0.1cm}
CS1&20&$28.3^{+3.8}_{-4.4}$&$37.8^{+10.7}_{-1.6}$& $43.5^{+3.1}_{-3.0}$ & $39.1^{+1.9}_{-7.5}$ & $39.3^{+0.3}_{-15.8}$ \\\vspace{0.1cm}
CS2&14& $27.0^{+15.0}_{-3.1}$ & $36.1^{+30.2}_{-8.9}$ & $44.6^{+3.4}_{-7.8}$ & $42.0^{+4.2}_{-6.8}$ & $37.9^{+13.0}_{-12.0}$ \\
Sample&21& $29.4^{+1.5}_{-4.8}$ & $38.5^{+8.2}_{-3.0}$ & $43.5^{+3.3}_{-5.3}$ & $38.5^{+1.6}_{-8.0}$ & $39.3^{+0.3}_{-14.6}$ \\
\hline\hline
\end{tabular}
\end{table*}

\begin{figure}
\begin{center}
\includegraphics[width=0.49\textwidth]{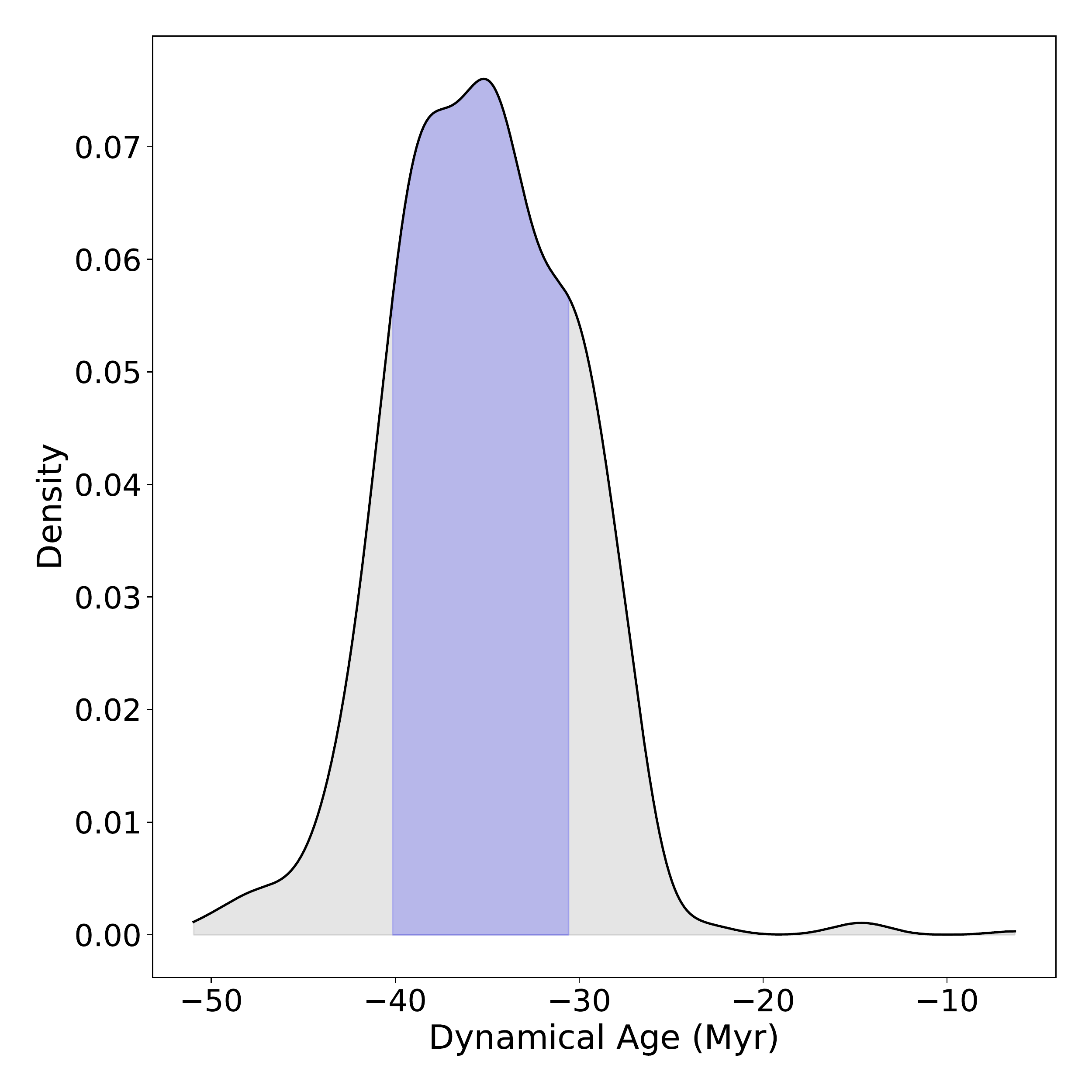}
\caption{Distribution of dynamical ages obtained after 1\,000 bootstrap repetitions with our final sample of 21 stars using the $S_{TCM}$ size estimator. The blue shaded region indicates the 68\% highest-density interval of our solution. The stellar orbits were integrated with the MWPotential2014 potential. }
\label{fig_age_distribution}
\end{center}
\end{figure}

We report the dynamical age of $38.5^{+1.6}_{-8.0}$~Myr obtained from the $S_{TCM}$ estimator as our final result for the reasons explained above. The asymmetric uncertainties of our solution result from the distribution of dynamical ages obtained from the bootstrap repetitions as illustrated in Figure~\ref{fig_age_distribution}. The bootstrap procedure employed in our analysis roughly evaluates the effect of sample completeness in our solution by computing the dynamical age of the association with different subsets of cluster members. Another possibility to investigate the uncertainty of the dynamical age consists in resampling the stellar positions, proper motions, parallaxes and radial velocities from a Gaussian multivariate distribution to evaluate the effect of the observational uncertainties and covariances in our solution. We generated 1\,000 samples using this resampling procedure, re-computed the 3D positions and 3D velocities of the stars, and derived the dynamical age for each sample from a traceback analysis as described before. Doing so, we obtain the age of 35.0$^{+5.3}_{-2.7}$~Myr which is fully consistent with the solution derived from the bootstrap procedure. Thus, we conclude that both methods predict a dynamical age for Tuc-Hor between 32.3 and 40.1~Myr within $1\sigma$ of the reported uncertainties. For consistency with other studies of this series performed by our team \citep[see][]{Miret-Roig2020} we report the solution obtained from the bootstrap procedure as our final result.

To further investigate the robustness of this solution we have reproduced the traceback analysis using a different model for the Galactic potential. We recomputed the stellar orbits using the axisymmetric potential of \citet{McMillan2017} and followed the same steps of our analysis performed with the MWPotential2014 Galactic potential. Doing so, we derived the dynamical age of $39.8^{+1.6}_{-8.9}$~Myr from the \citet{McMillan2017} potential which confirms our result obtained with the MWPotential2014 potential. This finding supports the conclusion obtained by \citet{Miret-Roig2020} that the changes in the dynamical age of the association due to the choice of the Galactic potential are minimal and consistent within the reported uncertainties.

\section{Discussion}\label{section4}

Let us now compare the dynamical age of Tuc-Hor obtained in this study with other results from the literature. \citet{Miret-Roig2018} derived the dynamical age of $5^{+23}_{-0}$~Myr using a similar method employed in this paper. Surprisingly, this was the only young stellar group investigated in that study for which the authors report a dynamical age that is smaller than other age estimates given in the literature. As explained below, this is not the case of the dynamical age derived by ourselves which is consistent with other age results obtained from different methods. The more precise astrometry and radial velocity information that is currently available for our study is undoubtedly one reason to explain the different results with respect to \citet{Miret-Roig2018}. This combined with a more robust sample selection and orbital analysis of individual stars has produced a clean sample of cluster members to estimate the size of the association. 

The recent study conducted by \citet{Kerr2022} expanded the census of young stars around the $\chi^{1}$ Fornacis cluster and investigated its traceback history with a more extensive complex of young stellar groups which they refer to as the ``Austral Complex'' and includes the Tuc-Hor, Carina and Columba associations. The authors derived the dynamical age of $27.2\pm5.9$~Myr for Tuc-Hor, but as explained in Section~4.2.5 of that study, they mention that their dynamical ages are likely not reliable for all subgroups of the Austral Complex. The reported age is consistent with the results obtained in this paper from the $S_{X}$ size estimator (see Table~\ref{tab_age}) which takes into account the dispersion of the stars in only one direction to estimate the size of the association. In addition to the different sample of stars, selection criteria and source of radial velocity data the methodology employed by \citet{Kerr2022} to infer the dynamical age is also different from the one used in our study. The authors estimate the size of the association by computing the median mutual distance between the stars in the sample for each time step, then obtain the dynamical age from the minimum of the curve. As illustrated in Figure~8 of that study this method produces a curve of the size of the association as function of time that exhibits a clear minimum e.g. in the case of the Carina and Columba associations. However, for the Tuc-Hor association we observe a plateau from about $t=-40$ to $t=-20$~Myr with no clear minimum of the curve similarly to what is observed in our analysis using the $S_{DCM}$ size estimator (see Figure~\ref{fig_size_time}). As explained before (see Sect.~\ref{section3}), the methodology that is used to estimate the size of the association is one critical point in any traceback analysis to derive dynamical ages and this is likely to be the main reason for the different result obtained in the two studies. It is also interesting to note that the minimum size of the association illustrated in Figure~8 of \citet{Kerr2022} is about twice as large as the minimum size of the association obtained from our analysis (see Figure~\ref{fig_size_time}). 

\begin{figure}
\begin{center}
\includegraphics[width=0.49\textwidth]{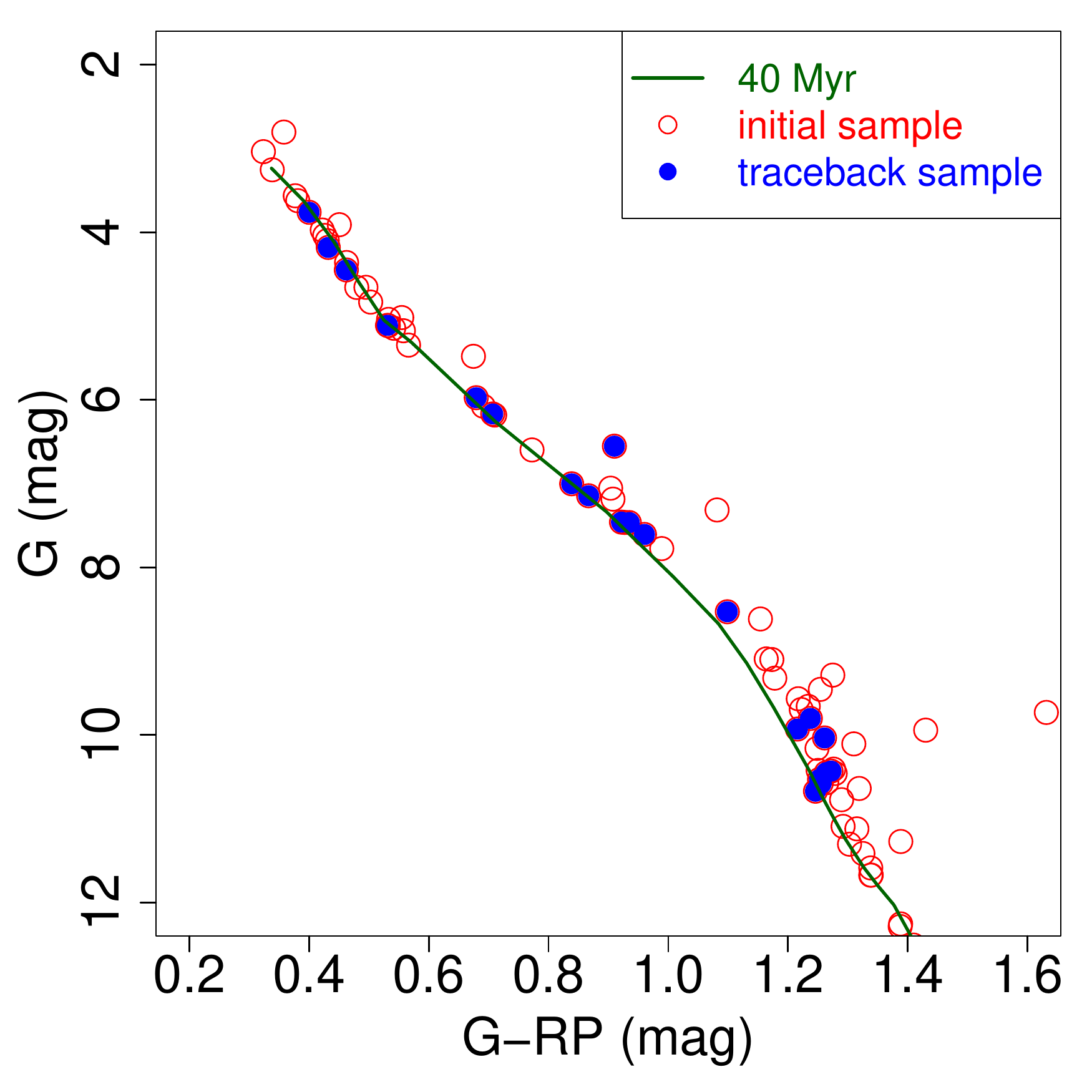}
\caption{Absolute colour magnitude diagram for Tuc-Hor stars. Open symbols indicate the stars from the initial literature sample (see Sect.~\ref{section2.1}) and filled symbols mark the 21 stars used in the traceback analysis. The solid line indicates the 40~Myr isochrone from the \citet{Baraffe2015} models.}
\label{fig_cmd}
\end{center}
\end{figure}

In addition to the traceback analysis, other methods have also been employed to investigate the age of the Tuc-Hor association. Early studies after recognition of the Tuc-Hor association as a nearby young stellar group tentatively assigned the age of about 30-40~Myr based on isochrones and the strength of H$\alpha$ emission lines of the stars \citep[see e.g.][]{Torres2000,Zuckerman2000,Zuckerman2004}. This age estimate was later confirmed by \citet{Kraus2014} with the result of 30~Myr based on isochrones using the sequence of temperatures for young stars given by \citet{Pecaut2013}. The subsequent study conducted by \citet{Bell2015} derived a somewhat older age of $45\pm4$~Myr from a semi-empirical isochrone fitting procedure based on different pre-main sequence star evolutionary models. The most recent results obtained by \citet{Kerr2022} indeed show a broad consistency with an age in the range of 30-50~Myr for Tuc-Hor using different age dating methods. The dynamical age of the association derived by ourselves in this study is nonetheless consistent with the various age results obtained from isochrones over the past years. In Figure~\ref{fig_cmd} we illustrate that the sample of stars used in the traceback analysis defines a narrow sequence in the color absolute magnitude diagram and coincides well with the 40~Myr isochrone derived from the \citet{Baraffe2015} models.

Lithium depletion in mid M dwarfs is another independent method to accurately derive the ages of young stellar groups \citep[see e.g.][]{Stauffer1998,Barrado1999,Binks2014,Galindo-Guil2022}. In this context, \citet{Mentuch2008} used the lithium depletion boundary (LDB) method to infer the age of the Tuc-Hor association and they obtained the result of $27\pm11$~Myr. However, the poor coverage of stars in their sample (in particular, M dwarfs) close to the location of the LDB makes their age estimate less accurate. The more robust study conducted by \citet{Kraus2014} based on the LDB method derived the age of $41\pm2$~Myr and $38\pm2$~Myr using the \citet{Baraffe1998} and \citet{Dantona1997} models, respectively. \citet{Galindo-Guil2022} derived LDB ages ranging from 41.3 to 51~Myr for Tuc-Hor using different isochrones. These numbers are in very good agreement with the dynamical age of $38.5^{+1.6}_{-8.0}$ obtained in this paper and support our result for the age of the Tuc-Hor association. \citet{Miret-Roig2020} reached a similar conclusion for the $\beta$~Pic moving group using the same methodology employed in this paper to derive the dynamical age of the association. It is therefore reassuring to find that the dynamical ages of these two young stellar groups, which are independent of evolutionary models, are consistent with the LDB ages. This finding also confirms, for the first time, the feasibility of deriving reliable dynamical ages of $\sim40$~Myr old stellar groups with the traceback technique. Altogether, these results suggest that the Tuc-Hor association is roughly twice as old as the $\beta$~Pic moving group \citep[$t\sim18.5^{+2.0}_{-2.4}$, ][]{Miret-Roig2020}. It is also remarkable to note that the minimum size of the association inferred from the $S_{TCM}$ estimator in both cases is of a few pc only (about 4~pc for Tuc-Hor and 7~pc for the $\beta$~Pic). The scatter (i.e. standard deviation) of the stellar positions for Tuc-Hor in the \textit{XYZ} direction at $t=-38.5$~Myr is 6.9, 9.6, and 3.8~pc which roughly constrains the size of a stellar association at birth time.

\section{Conclusions}\label{section5}

In this paper we derived the dynamical age of the Tuc-Hor association using the most precise astrometric data from the Gaia-DR3 catalogue and radial velocity information that is currently available. As part of our strategy, we measured the radial velocity for 40~stars by combining high-resolution spectra from public archives with our own observations performed with the UVES spectrograph. Our observations targeted the faintest sources in the sample with missing radial velocity information in the literature. The radial velocities derived by ourselves  are in good agreement with published results for the sources in common with other projects and they are often more precise. Doing so, we increased the number of stars with radial velocity data and delivered a homogeneous sample of precise measurements to study the kinematic properties of the Tuc-Hor association.

We performed a careful selection of the most likely kinematic members of the Tuc-Hor association based on the 3D space motion of the stars and the convergence of the stellar orbits back in time. Then, we conducted an extensive analysis to investigate the dynamical age of the association using a combination of different samples of stars, metrics to define the minimum size of the association, and models for the Galactic potential to trace the stellar orbits. We confirmed that the dynamical ages obtained from the different data sets explored in our analysis agree within 3~Myr using the same size estimator. Our analysis also revealed that the definition of the size of the association based on the trace of the covariance matrix returns a more robust estimate of the dynamical age of the Tuc-Hor association and that the differences in the dynamical age due to the choice of the Galactic potential that is employed in the traceback analysis are minimal. Our analysis demonstrates the need of a clean sample of cluster members, very accurate data and a robust statistical approach to derive reliable age results with the traceback method.

The dynamical age of $38.5^{+1.6}_{-8.0}$~Myr derived in this paper for the Tuc-Hor association is fully consistent with the various age estimates reported in the literature from isochrones. More importantly, our result reconciles, for the first time, the dynamical age of Tuc-Hor with the age of about 40~Myr obtained from the LDB method. A similar conclusion was reached by \citet{Miret-Roig2020} for the $\beta$~Pic moving group (dynamical age of $18.5^{+2.0}_{-2.4}$~Myr) using the same methodology used in this study to investigate the dynamical age of the Tuc-Hor association. Our results also illustrate the feasibility of deriving reliable dynamical ages for $\sim40$~Myr old stellar groups based on the traceback method. These findings represent the first step towards constructing a self-consistent age scale independent of stellar models for the young stellar groups in the Solar neighbourhood based on the 3D space motion of the stars.

\section*{Acknowledgements}
We thank the referee for constructive criticism that helped us to improve the manuscript. P.A.B.G. acknowledges financial support from the São Paulo Research Foundation (FAPESP) under grants 2020/12518-8 and 2021/11778-9. J.O. acknowledges financial support from ``Ayudas para contratos postdoctorales de investigación UNED 2021''. D.B. has been funded by Spanish MCIN/AEI/10.13039/501100011033 grant PID2019-107061GB-C61. This research has received funding from the European Research Council (ERC) under the European Union’s Horizon 2020 research and innovation programme (grant agreement No 682903, P.I. H. Bouy), and from the French State in the framework of the  ``Investments for the future” Program, IdEx Bordeaux, reference ANR-10-IDEX-03-02. This research has made use of the SIMBAD database, operated at CDS, Strasbourg, France. This work has made use of data from the European Space Agency (ESA) mission {\it Gaia} (\url{https://www.cosmos.esa.int/gaia}), processed by the {\it Gaia} Data Processing and Analysis Consortium (DPAC, \url{https://www.cosmos.esa.int/web/gaia/dpac/consortium}). Funding for the DPAC has been provided by national institutions, in particular the institutions participating in the {\it Gaia} Multilateral Agreement.

\section*{Data availability}
The data underlying this article are available in the article and in its online supplementary material.

\bibliographystyle{mnras}
\bibliography{references} 

\begin{thebibliography}{}
\makeatletter
\relax
\def\mn@urlcharsother{\let\do\@makeother \do\$\do\&\do\#\do\^\do\_\do\%\do\~}
\def\mn@doi{\begingroup\mn@urlcharsother \@ifnextchar [ {\mn@doi@}
  {\mn@doi@[]}}
\def\mn@doi@[#1]#2{\def\@tempa{#1}\ifx\@tempa\@empty \href
  {http://dx.doi.org/#2} {doi:#2}\else \href {http://dx.doi.org/#2} {#1}\fi
  \endgroup}
\def\mn@eprint#1#2{\mn@eprint@#1:#2::\@nil}
\def\mn@eprint@arXiv#1{\href {http://arxiv.org/abs/#1} {{\tt arXiv:#1}}}
\def\mn@eprint@dblp#1{\href {http://dblp.uni-trier.de/rec/bibtex/#1.xml}
  {dblp:#1}}
\def\mn@eprint@#1:#2:#3:#4\@nil{\def\@tempa {#1}\def\@tempb {#2}\def\@tempc
  {#3}\ifx \@tempc \@empty \let \@tempc \@tempb \let \@tempb \@tempa \fi \ifx
  \@tempb \@empty \def\@tempb {arXiv}\fi \@ifundefined
  {mn@eprint@\@tempb}{\@tempb:\@tempc}{\expandafter \expandafter \csname
  mn@eprint@\@tempb\endcsname \expandafter{\@tempc}}}

\bibitem[\protect\citeauthoryear{{Baraffe}, {Chabrier}, {Allard}  \&
  {Hauschildt}}{{Baraffe} et~al.}{1998}]{Baraffe1998}
{Baraffe} I.,  {Chabrier} G.,  {Allard} F.,   {Hauschildt} P.~H.,  1998, \aap,
  \href {https://ui.adsabs.harvard.edu/abs/1998A&A...337..403B} {337, 403}

\bibitem[\protect\citeauthoryear{{Baraffe}, {Homeier}, {Allard}  \&
  {Chabrier}}{{Baraffe} et~al.}{2015}]{Baraffe2015}
{Baraffe} I.,  {Homeier} D.,  {Allard} F.,   {Chabrier} G.,  2015, \mn@doi
  [\aap] {10.1051/0004-6361/201425481}, \href
  {https://ui.adsabs.harvard.edu/abs/2015A&A...577A..42B} {577, A42}

\bibitem[\protect\citeauthoryear{{Barrado y Navascu{\'e}s}, {Stauffer}  \&
  {Patten}}{{Barrado y Navascu{\'e}s} et~al.}{1999}]{Barrado1999}
{Barrado y Navascu{\'e}s} D.,  {Stauffer} J.~R.,   {Patten} B.~M.,  1999,
  \mn@doi [\apjl] {10.1086/312212}, \href
  {https://ui.adsabs.harvard.edu/abs/1999ApJ...522L..53B} {522, L53}

\bibitem[\protect\citeauthoryear{{Bell}, {Mamajek}  \& {Naylor}}{{Bell}
  et~al.}{2015}]{Bell2015}
{Bell} C. P.~M.,  {Mamajek} E.~E.,   {Naylor} T.,  2015, \mn@doi [\mnras]
  {10.1093/mnras/stv1981}, \href
  {https://ui.adsabs.harvard.edu/abs/2015MNRAS.454..593B} {454, 593}

\bibitem[\protect\citeauthoryear{{Binks} \& {Jeffries}}{{Binks} \&
  {Jeffries}}{2014}]{Binks2014}
{Binks} A.~S.,  {Jeffries} R.~D.,  2014, \mn@doi [\mnras]
  {10.1093/mnrasl/slt141}, \href
  {https://ui.adsabs.harvard.edu/abs/2014MNRAS.438L..11B} {438, L11}

\bibitem[\protect\citeauthoryear{{Blanco-Cuaresma}}{{Blanco-Cuaresma}}{2019}]{BlancoCuaresma2019}
{Blanco-Cuaresma} S.,  2019, \mn@doi [\mnras] {10.1093/mnras/stz549}, \href
  {https://ui.adsabs.harvard.edu/abs/2019MNRAS.486.2075B} {486, 2075}

\bibitem[\protect\citeauthoryear{{Blanco-Cuaresma}, {Soubiran}, {Heiter}  \&
  {Jofr{\'e}}}{{Blanco-Cuaresma} et~al.}{2014}]{BlancoCuaresma2014}
{Blanco-Cuaresma} S.,  {Soubiran} C.,  {Heiter} U.,   {Jofr{\'e}} P.,  2014,
  \mn@doi [\aap] {10.1051/0004-6361/201423945}, \href
  {https://ui.adsabs.harvard.edu/abs/2014A&A...569A.111B} {569, A111}

\bibitem[\protect\citeauthoryear{{Boccaletti}, {Augereau}, {Baudoz}, {Pantin}
  \& {Lagrange}}{{Boccaletti} et~al.}{2009}]{Boccaletti2009}
{Boccaletti} A.,  {Augereau} J.~C.,  {Baudoz} P.,  {Pantin} E.,   {Lagrange}
  A.~M.,  2009, \mn@doi [\aap] {10.1051/0004-6361:200811067}, \href
  {https://ui.adsabs.harvard.edu/abs/2009A&A...495..523B} {495, 523}

\bibitem[\protect\citeauthoryear{{Bovy}}{{Bovy}}{2015}]{Bovy2015}
{Bovy} J.,  2015, \mn@doi [\apjs] {10.1088/0067-0049/216/2/29}, \href
  {https://ui.adsabs.harvard.edu/abs/2015ApJS..216...29B} {216, 29}

\bibitem[\protect\citeauthoryear{{Chauvin}, {Lagrange}, {Dumas}, {Zuckerman},
  {Mouillet}, {Song}, {Beuzit}  \& {Lowrance}}{{Chauvin}
  et~al.}{2004}]{Chauvin2004}
{Chauvin} G.,  {Lagrange} A.~M.,  {Dumas} C.,  {Zuckerman} B.,  {Mouillet} D.,
  {Song} I.,  {Beuzit} J.~L.,   {Lowrance} P.,  2004, \mn@doi [\aap]
  {10.1051/0004-6361:200400056}, \href
  {https://ui.adsabs.harvard.edu/abs/2004A&A...425L..29C} {425, L29}

\bibitem[\protect\citeauthoryear{{D'Antona} \& {Mazzitelli}}{{D'Antona} \&
  {Mazzitelli}}{1997}]{Dantona1997}
{D'Antona} F.,  {Mazzitelli} I.,  1997, \memsai, \href
  {https://ui.adsabs.harvard.edu/abs/1997MmSAI..68..807D} {68, 807}

\bibitem[\protect\citeauthoryear{{Dekker}, {D'Odorico}, {Kaufer}, {Delabre}  \&
  {Kotzlowski}}{{Dekker} et~al.}{2000}]{Dekker2000}
{Dekker} H.,  {D'Odorico} S.,  {Kaufer} A.,  {Delabre} B.,   {Kotzlowski} H.,
  2000, in {Iye} M.,  {Moorwood} A.~F.,  eds,  Society of Photo-Optical
  Instrumentation Engineers (SPIE) Conference Series Vol. 4008, Optical and IR
  Telescope Instrumentation and Detectors. pp 534--545,
  \mn@doi{10.1117/12.395512}

\bibitem[\protect\citeauthoryear{{Ducourant}, {Teixeira}, {Galli}, {Le
  Campion}, {Krone-Martins}, {Zuckerman}, {Chauvin}  \& {Song}}{{Ducourant}
  et~al.}{2014}]{Ducourant2014}
{Ducourant} C.,  {Teixeira} R.,  {Galli} P.~A.~B.,  {Le Campion} J.~F.,
  {Krone-Martins} A.,  {Zuckerman} B.,  {Chauvin} G.,   {Song} I.,  2014,
  \mn@doi [\aap] {10.1051/0004-6361/201322075}, \href
  {https://ui.adsabs.harvard.edu/abs/2014A&A...563A.121D} {563, A121}

\bibitem[\protect\citeauthoryear{{Gagn{\'e}} \& {Faherty}}{{Gagn{\'e}} \&
  {Faherty}}{2018}]{Gagne2018c}
{Gagn{\'e}} J.,  {Faherty} J.~K.,  2018, \mn@doi [\apj]
  {10.3847/1538-4357/aaca2e}, \href
  {https://ui.adsabs.harvard.edu/abs/2018ApJ...862..138G} {862, 138}

\bibitem[\protect\citeauthoryear{{Gagn{\'e}} et~al.,}{{Gagn{\'e}}
  et~al.}{2018a}]{Gagne2018a}
{Gagn{\'e}} J.,  et~al., 2018a, \mn@doi [\apj] {10.3847/1538-4357/aaae09},
  \href {https://ui.adsabs.harvard.edu/abs/2018ApJ...856...23G} {856, 23}

\bibitem[\protect\citeauthoryear{{Gagn{\'e}}, {Roy-Loubier}, {Faherty}, {Doyon}
   \& {Malo}}{{Gagn{\'e}} et~al.}{2018b}]{Gagne2018b}
{Gagn{\'e}} J.,  {Roy-Loubier} O.,  {Faherty} J.~K.,  {Doyon} R.,   {Malo} L.,
  2018b, \mn@doi [\apj] {10.3847/1538-4357/aac2b8}, \href
  {https://ui.adsabs.harvard.edu/abs/2018ApJ...860...43G} {860, 43}

\bibitem[\protect\citeauthoryear{{Gaia Collaboration} et~al.,}{{Gaia
  Collaboration} et~al.}{2016}]{Gaia}
{Gaia Collaboration} et~al., 2016, \mn@doi [\aap]
  {10.1051/0004-6361/201629272}, \href
  {https://ui.adsabs.harvard.edu/abs/2016A&A...595A...1G} {595, A1}

\bibitem[\protect\citeauthoryear{{Gaia Collaboration} et~al.,}{{Gaia
  Collaboration} et~al.}{2022}]{GaiaDR3}
{Gaia Collaboration} et~al., 2022, arXiv e-prints, \href
  {https://ui.adsabs.harvard.edu/abs/2022arXiv220800211G} {p. arXiv:2208.00211}

\bibitem[\protect\citeauthoryear{{Galindo-Guil} et~al.,}{{Galindo-Guil}
  et~al.}{2022}]{Galindo-Guil2022}
{Galindo-Guil} F.~J.,  et~al., 2022, \mn@doi [\aap]
  {10.1051/0004-6361/202141114}, \href
  {https://ui.adsabs.harvard.edu/abs/2022A&A...664A..70G} {664, A70}

\bibitem[\protect\citeauthoryear{{Golimowski} et~al.,}{{Golimowski}
  et~al.}{2006}]{Golimowski2006}
{Golimowski} D.~A.,  et~al., 2006, \mn@doi [\aj] {10.1086/503801}, \href
  {https://ui.adsabs.harvard.edu/abs/2006AJ....131.3109G} {131, 3109}

\bibitem[\protect\citeauthoryear{{Gontcharov}}{{Gontcharov}}{2006}]{Gontcharov2006}
{Gontcharov} G.~A.,  2006, \mn@doi [Astronomy Letters]
  {10.1134/S1063773706110065}, \href
  {https://ui.adsabs.harvard.edu/abs/2006AstL...32..759G} {32, 759}

\bibitem[\protect\citeauthoryear{{Holmberg}, {Nordstr{\"o}m}  \&
  {Andersen}}{{Holmberg} et~al.}{2007}]{Holmberg2007}
{Holmberg} J.,  {Nordstr{\"o}m} B.,   {Andersen} J.,  2007, \mn@doi [\aap]
  {10.1051/0004-6361:20077221}, \href
  {https://ui.adsabs.harvard.edu/abs/2007A&A...475..519H} {475, 519}

\bibitem[\protect\citeauthoryear{{Johnson} \& {Soderblom}}{{Johnson} \&
  {Soderblom}}{1987}]{Johnson1987}
{Johnson} D. R.~H.,  {Soderblom} D.~R.,  1987, \mn@doi [\aj] {10.1086/114370},
  \href {https://ui.adsabs.harvard.edu/abs/1987AJ.....93..864J} {93, 864}

\bibitem[\protect\citeauthoryear{{Kerr}, {Kraus}, {Murphy}, {Krolikowski},
  {Bedding}  \& {Rizzuto}}{{Kerr} et~al.}{2022}]{Kerr2022}
{Kerr} R.,  {Kraus} A.~L.,  {Murphy} S.~J.,  {Krolikowski} D.~M.,  {Bedding}
  T.~R.,   {Rizzuto} A.~C.,  2022, arXiv e-prints, \href
  {https://ui.adsabs.harvard.edu/abs/2022arXiv221100123K} {p. arXiv:2211.00123}

\bibitem[\protect\citeauthoryear{{Kraus}, {Shkolnik}, {Allers}  \&
  {Liu}}{{Kraus} et~al.}{2014}]{Kraus2014}
{Kraus} A.~L.,  {Shkolnik} E.~L.,  {Allers} K.~N.,   {Liu} M.~C.,  2014,
  \mn@doi [\aj] {10.1088/0004-6256/147/6/146}, \href
  {https://ui.adsabs.harvard.edu/abs/2014AJ....147..146K} {147, 146}

\bibitem[\protect\citeauthoryear{{Lagrange} et~al.,}{{Lagrange}
  et~al.}{2010}]{Lagrange2010}
{Lagrange} A.~M.,  et~al., 2010, \mn@doi [Science] {10.1126/science.1187187},
  \href {https://ui.adsabs.harvard.edu/abs/2010Sci...329...57L} {329, 57}

\bibitem[\protect\citeauthoryear{{McMillan}}{{McMillan}}{2017}]{McMillan2017}
{McMillan} P.~J.,  2017, \mn@doi [\mnras] {10.1093/mnras/stw2759}, \href
  {https://ui.adsabs.harvard.edu/abs/2017MNRAS.465...76M} {465, 76}

\bibitem[\protect\citeauthoryear{{Mentuch}, {Brandeker}, {van Kerkwijk},
  {Jayawardhana}  \& {Hauschildt}}{{Mentuch} et~al.}{2008}]{Mentuch2008}
{Mentuch} E.,  {Brandeker} A.,  {van Kerkwijk} M.~H.,  {Jayawardhana} R.,
  {Hauschildt} P.~H.,  2008, \mn@doi [\apj] {10.1086/592764}, \href
  {https://ui.adsabs.harvard.edu/abs/2008ApJ...689.1127M} {689, 1127}

\bibitem[\protect\citeauthoryear{{Miret-Roig}, {Antoja}, {Romero-G{\'o}mez}  \&
  {Figueras}}{{Miret-Roig} et~al.}{2018}]{Miret-Roig2018}
{Miret-Roig} N.,  {Antoja} T.,  {Romero-G{\'o}mez} M.,   {Figueras} F.,  2018,
  \mn@doi [\aap] {10.1051/0004-6361/201731976}, \href
  {https://ui.adsabs.harvard.edu/abs/2018A&A...615A..51M} {615, A51}

\bibitem[\protect\citeauthoryear{{Miret-Roig} et~al.,}{{Miret-Roig}
  et~al.}{2020}]{Miret-Roig2020}
{Miret-Roig} N.,  et~al., 2020, \mn@doi [\aap] {10.1051/0004-6361/202038765},
  \href {https://ui.adsabs.harvard.edu/abs/2020A&A...642A.179M} {642, A179}

\bibitem[\protect\citeauthoryear{{Newton} et~al.,}{{Newton}
  et~al.}{2019}]{Newton2019}
{Newton} E.~R.,  et~al., 2019, \mn@doi [\apjl] {10.3847/2041-8213/ab2988},
  \href {https://ui.adsabs.harvard.edu/abs/2019ApJ...880L..17N} {880, L17}

\bibitem[\protect\citeauthoryear{{Pecaut} \& {Mamajek}}{{Pecaut} \&
  {Mamajek}}{2013}]{Pecaut2013}
{Pecaut} M.~J.,  {Mamajek} E.~E.,  2013, \mn@doi [\apjs]
  {10.1088/0067-0049/208/1/9}, \href
  {https://ui.adsabs.harvard.edu/abs/2013ApJS..208....9P} {208, 9}

\bibitem[\protect\citeauthoryear{Pedregosa et~al.,}{Pedregosa
  et~al.}{2011}]{scikit-learn}
Pedregosa F.,  et~al., 2011, Journal of Machine Learning Research, 12, 2825

\bibitem[\protect\citeauthoryear{Rousseeuw \& Driessen}{Rousseeuw \&
  Driessen}{1999}]{Rousseeuw1999}
Rousseeuw P.~J.,  Driessen K.~V.,  1999, \mn@doi [Technometrics]
  {10.1080/00401706.1999.10485670}, 41, 212

\bibitem[\protect\citeauthoryear{{Sch{\"o}nrich}, {Binney}  \&
  {Dehnen}}{{Sch{\"o}nrich} et~al.}{2010}]{Schoenrich2010}
{Sch{\"o}nrich} R.,  {Binney} J.,   {Dehnen} W.,  2010, \mn@doi [\mnras]
  {10.1111/j.1365-2966.2010.16253.x}, \href
  {https://ui.adsabs.harvard.edu/abs/2010MNRAS.403.1829S} {403, 1829}

\bibitem[\protect\citeauthoryear{{Smith} \& {Terrile}}{{Smith} \&
  {Terrile}}{1984}]{Smith1984}
{Smith} B.~A.,  {Terrile} R.~J.,  1984, \mn@doi [Science]
  {10.1126/science.226.4681.1421}, \href
  {https://ui.adsabs.harvard.edu/abs/1984Sci...226.1421S} {226, 1421}

\bibitem[\protect\citeauthoryear{{Soderblom}}{{Soderblom}}{2010}]{Soderblom2010}
{Soderblom} D.~R.,  2010, \mn@doi [\araa]
  {10.1146/annurev-astro-081309-130806}, \href
  {https://ui.adsabs.harvard.edu/abs/2010ARA&A..48..581S} {48, 581}

\bibitem[\protect\citeauthoryear{{Stauffer}, {Schultz}  \&
  {Kirkpatrick}}{{Stauffer} et~al.}{1998}]{Stauffer1998}
{Stauffer} J.~R.,  {Schultz} G.,   {Kirkpatrick} J.~D.,  1998, \mn@doi [\apjl]
  {10.1086/311379}, \href
  {https://ui.adsabs.harvard.edu/abs/1998ApJ...499L.199S} {499, L199}

\bibitem[\protect\citeauthoryear{{Torres}, {da Silva}, {Quast}, {de la Reza}
  \& {Jilinski}}{{Torres} et~al.}{2000}]{Torres2000}
{Torres} C. A.~O.,  {da Silva} L.,  {Quast} G.~R.,  {de la Reza} R.,
  {Jilinski} E.,  2000, \mn@doi [\aj] {10.1086/301539}, \href
  {https://ui.adsabs.harvard.edu/abs/2000AJ....120.1410T} {120, 1410}

\bibitem[\protect\citeauthoryear{{Torres}, {Quast}, {da Silva}, {de La Reza},
  {Melo}  \& {Sterzik}}{{Torres} et~al.}{2006}]{Torres2006}
{Torres} C.~A.~O.,  {Quast} G.~R.,  {da Silva} L.,  {de La Reza} R.,  {Melo}
  C.~H.~F.,   {Sterzik} M.,  2006, \mn@doi [\aap] {10.1051/0004-6361:20065602},
  \href {https://ui.adsabs.harvard.edu/abs/2006A&A...460..695T} {460, 695}

\bibitem[\protect\citeauthoryear{{Torres}, {Quast}, {Melo}  \&
  {Sterzik}}{{Torres} et~al.}{2008}]{Torres2008}
{Torres} C.~A.~O.,  {Quast} G.~R.,  {Melo} C.~H.~F.,   {Sterzik} M.~F.,  2008,
  in {Reipurth} B.,  ed., , Vol.~5, Handbook of Star Forming Regions, Volume
  II.
p.~757

\bibitem[\protect\citeauthoryear{{Zucker}}{{Zucker}}{2003}]{Zucker2003}
{Zucker} S.,  2003, \mn@doi [\mnras] {10.1046/j.1365-8711.2003.06633.x}, \href
  {https://ui.adsabs.harvard.edu/abs/2003MNRAS.342.1291Z} {342, 1291}

\bibitem[\protect\citeauthoryear{{Zuckerman} \& {Song}}{{Zuckerman} \&
  {Song}}{2004}]{Zuckerman2004}
{Zuckerman} B.,  {Song} I.,  2004, \mn@doi [\araa]
  {10.1146/annurev.astro.42.053102.134111}, \href
  {https://ui.adsabs.harvard.edu/abs/2004ARA&A..42..685Z} {42, 685}

\bibitem[\protect\citeauthoryear{{Zuckerman} \& {Webb}}{{Zuckerman} \&
  {Webb}}{2000}]{Zuckerman2000}
{Zuckerman} B.,  {Webb} R.~A.,  2000, \mn@doi [\apj] {10.1086/308897}, \href
  {https://ui.adsabs.harvard.edu/abs/2000ApJ...535..959Z} {535, 959}

\bibitem[\protect\citeauthoryear{{Zuckerman}, {Song}  \& {Webb}}{{Zuckerman}
  et~al.}{2001}]{Zuckerman2001}
{Zuckerman} B.,  {Song} I.,   {Webb} R.~A.,  2001, \mn@doi [\apj]
  {10.1086/322305}, \href
  {https://ui.adsabs.harvard.edu/abs/2001ApJ...559..388Z} {559, 388}

\bibitem[\protect\citeauthoryear{{de la Reza}, {Jilinski}  \& {Ortega}}{{de la
  Reza} et~al.}{2006}]{delaReza2006}
{de la Reza} R.,  {Jilinski} E.,   {Ortega} V.~G.,  2006, \mn@doi [\aj]
  {10.1086/501525}, \href
  {https://ui.adsabs.harvard.edu/abs/2006AJ....131.2609D} {131, 2609}

\makeatother
\end{thebibliography}

\appendix
\section{Tables}

\begin{table*}
\centering
\caption{Radial velocities derived in this paper. We provide for each star the Gaia-DR3 source identifier and coordinates, modified julian date (MJD), radial velocity (and its uncertainty), radial velocity scatter computed from three different masks (as explained in the text of Sect.~\ref{section2.3}), SNR of the spectrum, program identifier of the collected spectrum, instrument and multiplicity flag ('0' = single star, '1'= binary or multiple system). This table will be available in its entirety in machine-readable form.
\label{tab_RVs}}
\begin{tabular}{lccccccccc}
\hline\hline
Star identifier&$\alpha$&$\delta$&MJD&$RV$&$\sigma_{RV}$&SNR&Program ID&Instrument&Flag\\
&(deg)&(deg)&(days)&(km/s)&(km/s)&&&\\
\hline\hline

Gaia EDR3 4995853014646189696 & 1.46951961 & -41.7534081 & 52981.1 & $ 0.75 \pm 0.18 $& 0.46 & 172 & 072.A-9006 & FEROS & 0 \\
Gaia EDR3 4995853014646189696 & 1.46951961 & -41.7534081 & 54313.2 & $ 2.34 \pm 0.17 $& 0.14 & 90 & 079.A-9017 & FEROS & 0 \\
Gaia EDR3 4995853014646189696 & 1.46951961 & -41.7534081 & 53870.4 & $ 2.31 \pm 0.17 $& 0.14 & 205 & 077.C-0573 & FEROS & 0 \\
Gaia EDR3 4995853014646189696 & 1.46951961 & -41.7534081 & 53871.4 & $ 2.13 \pm 0.17 $& 0.12 & 197 & 077.C-0573 & FEROS & 0 \\
Gaia EDR3 4995853014646189696 & 1.46951961 & -41.7534081 & 53575.2 & $ 2.04 \pm 0.17 $& 0.12 & 327 & 075.A-9010 & FEROS & 0 \\
Gaia EDR3 4995853014646189696 & 1.46951961 & -41.7534081 & 57317.1 & $ 2.12 \pm 0.17 $& 0.12 & 224 & 094.A-9012 & FEROS & 0 \\
Gaia EDR3 4995853014646189696 & 1.46951961 & -41.7534081 & 54120.0 & $ 2.16 \pm 0.18 $& 0.12 & 203 & 078.C-0378 & FEROS & 0 \\
Gaia EDR3 4995853014646189696 & 1.46951961 & -41.7534081 & 53869.4 & $ 2.17 \pm 0.18 $& 0.12 & 222 & 077.C-0573 & FEROS & 0 \\
Gaia EDR3 4995853014646189696 & 1.46951961 & -41.7534081 & 53869.4 & $ 2.17 \pm 0.17 $& 0.13 & 209 & 077.C-0573 & FEROS & 0 \\
Gaia EDR3 4995853014646189696 & 1.46951961 & -41.7534081 & 54316.3 & $ 1.78 \pm 0.17 $& 0.13 & 316 & 079.A-9017 & FEROS & 0 \\

\hline\hline
\end{tabular}
\end{table*}

\begin{landscape}
\begin{table}
\centering
\caption{Properties of the 90 stars in Tuc-Hor from our initial input sample with available astrometric data in the Gaia-DR3 catalogue. We provide for each star the source identifier, position, proper motion, parallax, radial velocity from Gaia-DR3, radial velocity derived in this study and binary flag (``0'' = single star, ``1'' = binary or multiple star). The last three columns indicate whether the star is retained in our sample, CS1 and CS2 after applying the selection given in Sect.~\ref{section2.4} (``0'' = discarded, ``1'' = selected). This table will be available in its entirety in machine-readable form.
\label{tab_data}}
\begin{tabular}{lccccccccccc}
\hline\hline
Star identifier&$\alpha$&$\delta$&$\mu_{\alpha}\cos\delta$&$\mu_{\delta}$&$\varpi$&$RV$ (GaiaDR3)&$RV$ (This Paper)&Binary&Sample&CS1&CS2\\
&(deg)&(deg)&(mas/yr)&(mas/yr)&(mas)&(km/s)&(km/s)\\

\hline\hline

Gaia DR3 4995853014646189696 & 1.46951961 & -41.7534081 & $ 97.863 \pm 0.017 $& $ -76.502 \pm 0.018 $& $ 25.753 \pm 0.021 $& $ 2.05 \pm 0.16 $& $ 2.13 \pm 0.35 $& 0 & 1 & 1 & 1 \\
Gaia DR3 4688292536884945536 & 3.47227133 & -74.6885102 & $ 82.759 \pm 0.017 $& $ -49.159 \pm 0.015 $& $ 21.803 \pm 0.012 $& $ 9.09 \pm 0.19 $& $ 9.78 \pm 0.29 $& 1 & 0 & 0 & 0 \\
Gaia DR3 2320267025518037760 & 3.90352546 & -29.7671707 & $ 107.826 \pm 0.029 $& $ -79.918 \pm 0.025 $& $ 27.578 \pm 0.036 $& $ 0.51 \pm 1.28 $& $ -0.93 \pm 0.51 $& 0 & 1 & 1 & 0 \\
Gaia DR3 4901229043960053248 & 4.60974456 & -63.4777579 & $ 90.053 \pm 0.016 $& $ -59.207 \pm 0.018 $& $ 23.356 \pm 0.016 $& $ 6.53 \pm 0.16 $& $ 6.90 \pm 0.83 $& 0 & 1 & 1 & 1 \\
Gaia DR3 4902014095262270336 & 6.31191129 & -61.5136532 & $ 88.016 \pm 0.011 $& $ -56.164 \pm 0.012 $& $ 22.651 \pm 0.012 $& $ 6.21 \pm 0.38 $& $ 6.54 \pm 0.44 $& 0 & 1 & 1 & 1 \\
Gaia DR3 4997583955185946496 & 7.66554810 & -38.1660213 & $ 92.708 \pm 0.014 $& $ -63.488 \pm 0.019 $& $ 23.243 \pm 0.019 $& & $ 0.35 \pm 1.68 $& 0 & 1 & 1 & 0 \\
Gaia DR3 5000558409016727296 & 9.89783704 & -38.2885340 & $ 101.788 \pm 0.019 $& $ -66.587 \pm 0.028 $& $ 24.996 \pm 0.023 $& & $ 3.69 \pm 0.24 $& 0 & 1 & 1 & 0 \\
Gaia DR3 5014168610622763008 & 18.26564125 & -34.6387181 & $ 110.223 \pm 0.015 $& $ -60.217 \pm 0.015 $& $ 26.238 \pm 0.018 $& $ 6.88 \pm 37.00 $& $ 4.93 \pm 1.78 $& 0 & 1 & 1 & 0 \\
Gaia DR3 5015973252801591808 & 20.51906365 & -33.6179172 & $ 109.930 \pm 0.012 $& $ -57.349 \pm 0.015 $& $ 25.904 \pm 0.016 $& $ 4.50 \pm 0.21 $& $ 4.48 \pm 0.36 $& 0 & 1 & 1 & 1 \\
Gaia DR3 4691426694779173376 & 20.79815144 & -69.3606746 & $ 93.388 \pm 0.085 $& $ -26.499 \pm 0.081 $& $ 22.462 \pm 0.078 $& & & 0 & 0 & 0 & 0 \\

\hline\hline
\end{tabular}
\end{table}
\end{landscape}
\clearpage

\bsp	
\label{lastpage}
\end{document}